\documentclass[a4paper,11pt]{article}
\pdfoutput=1
\usepackage{jcappub}
\usepackage{slashed}
\usepackage{mathtools}
\usepackage[T1]{fontenc}
\usepackage[normalem]{ulem}

\usepackage[usenames,dvipsnames,svgnames]{xcolor}
\usepackage[numbers,sort&compress]{natbib}
\usepackage[utf8]{inputenc}
\usepackage{amsmath,amssymb,amsfonts,amsthm}
\usepackage{tabularx}
\usepackage{multirow}
\usepackage{dsfont}
\newcolumntype{Y}{>{\centering\arraybackslash}X}
\usepackage{xcolor}
\usepackage{url}
\usepackage[colorlinks=true]{hyperref}
\usepackage{graphicx}
\usepackage{cancel,slashed}
\usepackage[english]{babel}

\title{Probing inelastic sub-GeV dark matter at the DUNE near detector}
\author[a]{Amalia Betancur,}
\author[b]{Gustavo Castrillón,}
\author[b]{\'Oscar Zapata}

\affiliation[a]{Grupo F\'isica Te\'orica y Aplicada, Universidad EIA, A.A. 7516, Envigado, Colombia}
\affiliation[b]{Instituto de Física, Universidad de Antioquia\\
Calle 70 \# 52-21, Apartado Aéreo 1226, Medellín, Colombia}

\emailAdd{amalia.betancur@eia.edu.co}
\emailAdd{gadolfo.castrillon@udea.edu.co}
\emailAdd{oalberto.zapata@udea.edu.co}

\abstract{
We study inelastic dark matter (iDM) in a minimal and ultraviolet-complete framework in which the dark photon mass arises from a dark Higgs mechanism. The spontaneous breaking of a $U(1)_D$ symmetry splits the Dirac fermion into two Majorana states, thus opening new annihilation channels mediated by the dark Higgs. Focusing on sub-GeV dark matter, we assess the sensitivity of DUNE's liquid argon cube at the near detector (ND-LAr) to this scenario. We find that the ND-LAr can probe regions of parameter space consistent with the observed relic abundance due to these new annihilation channels, particularly for large dark photon-to-DM mass ratios where decay-based searches lose sensitivity. Our results highlight the complementarity between cosmological constraints and fixed-target experiments and demonstrate the potential of DUNE's ND-LAr to explore iDM scenarios with extended dark sectors.
}
\begin{document}
\begin{flushright}
\end{flushright}
\maketitle

%%%%%%%%%%%%%%%%%%%%%%%%%%%%%%%%%%%%%
\section{Introduction}\label{sec:intro}
%%%%%%%%%%%%%%%%%%%%%%%%%%%%%%%%%%%%%
Despite the overwhelming evidence of the existence of Dark Matter (DM)~\cite{Cirelli:2024ssz} and large efforts to understand its properties, the DM nature remains unknown. Weakly interacting massive particles (WIMPs) have been the most widely studied candidates~\cite{Jungman:1995df, Bertone:2004pz}. They naturally arise in beyond the Standard Model (SM) theories with the additional advantage of being thermally produced and allow for their detection through different search strategies---actually WIMPs have been the primary target of experimental searches for decades~\cite{Bertone:2016nfn,Roszkowski:2017nbc,Arcadi:2017kky}. However, the absence of signals at the electroweak scale from direct detection (DD)~\cite{LZ:2024zvo,PandaX:2024qfu,XENON:2025vwd} and collider experiments~\cite{CMS:2024zqs,ATLAS:2024fdw} has prompted the community to explore lighter DM candidates, for instance in the sub-GeV mass regime~\cite{Alexander:2016aln, Battaglieri:2017aum, Knapen:2017xzo, Lin:2019uvt}. 

A simple and well-motivated extension of the SM is obtained by enlarging its gauge symmetry with an additional $U(1)_{D}$ factor, whose mediator is the dark photon ($A^\prime$)~\cite{Okun:1982xi,Holdom:1985ag}. This dark-sector framework provides a minimal portal between DM and the SM~\cite{Bauer:2018onh,Fabbrichesi:2020wbt}, allowing the DM to interact with ordinary matter through the exchange of the dark photon in a wide variety of models featuring a fermion or scalar candidates~\cite{Berlin:2018bsc}. In the case of Dirac fermion thermally produced, with the fermion being lighter than the dark photon, annihilation into SM fermions is typically dominated by an $s$-wave process that remains unsuppressed at low velocities. As a result, the annihilation is still efficient at the time of recombination, placing the model in strong tension with CMB constraints~\cite{Slatyer:2015jla,Planck:2018vyg}.  Nevertheless, this constraint can be significantly relaxed if the Dirac fermion is split into two non-degenerate states~\cite{Izaguirre:2015yja,Izaguirre:2015zva}, thus realizing an inelastic DM (iDM) scenario~\cite{Tucker-Smith:2001myb}.

Although most of the works remain agnostic about the mechanism underlying the origin of both the dark photon mass and the fermion mass splitting, recent studies have started exploring the ultraviolet completion of the model and its phenomenological implications. In particular, when the dark photon acquires its mass through a Higgs mechanism via the spontaneous breaking of the $U(1)_{D}$ gauge symmetry by a new scalar field, commonly referred to as the dark Higgs~\cite{Patt:2006fw,Ahlers:2008qc,Batell:2009yf}, a mass splitting between the Majorana states arises naturally.  In this case, the mass splitting is intrinsically linked to the dark photon mass, implying that the dark Higgs plays an important role in the model's phenomenology.
Indeed, works such as Refs.~\cite{Darme:2018jmx,Duerr:2020muu,Fitzpatrick:2021cij,Darme:2021huc, Li:2021rzt, Ferber:2023iso, Reilly:2023frg, Garcia:2024uwf, Ko:2025drr, Liu:2025abt, Belle-II:2025bhd, DallaValleGarcia:2025cwf} have demonstrated that including the dark Higgs renders additional channels relevant for setting the DM relic abundance, while also opening new avenues for current experiment searches. This highlights the importance of  phenomenological studies of the complete model.

These dark sectors are often considered in the sub-GeV mass regime, which opens new opportunities for exploration beyond the more studied weak-scale DM. In this context, fixed-target experiments—including neutrino fixed-target facilities—have been shown to provide an ideal setting for probing sub-GeV DM~\cite{DeRomeri:2019kic,Breitbach:2021gvv,deNiverville:2018dbu,Berlin:2020uwy}, including iDM~\cite{Batell:2021ooj,NA64:2023wbi,Mongillo:2023hbs}. On the one hand, the energy ranges of the fixed-target collisions lie at the right scale for producing DM particles with sub-GeV masses. On the other hand, the detectors located downstream from the collision point offer a large target material where DM particles may interact with either electrons or nucleons, thus imprinting a signal that the precise sensors may detect. 
Several studies have begun investigating the potential to probe these DM models coupled with a dark photon, as explored in, e.g. Refs.~\cite{deNiverville:2018dbu,DeRomeri:2019kic,Breitbach:2021gvv,NOvA:2025ykl}. Moreover, current fixed-target experiments are using their accumulated luminosity to place constraints on these sectors. Recently, the COHERENT experiment~\cite{COHERENT:2021pvd} presented exclusion limits for scalar DM, NA64 presented exclusions for fermion and scalar DM~\cite{NA64:2023wbi}, while the authors of Ref.~\cite{Mongillo:2023hbs} used the NA64 results to constrain iDM model. All works show that their techniques can outperform DD experiments, despite having large backgrounds that may compete with DM signals~\cite{COHERENT:2021pvd}. 

In this work, we investigate the phenomenology of inelastic DM in the setup where the dark photon mass arises from a dark Higgs mechanism. We show that the presence of the dark Higgs qualitatively modifies the relic-abundance dynamics by opening forbidden and $p$-wave–suppressed annihilation channels that can dominate over the conventional inelastic process. This shifts the viable parameter space toward regions that are difficult to probe with traditional decay-based searches, particularly at large dark photon–to–DM mass ratios. We therefore assess the sensitivity of fixed-target experiments to this scenario and, for concreteness, focus on the forthcoming DUNE near detector liquid argon cube (ND-LAr)~\cite{DUNE:2021tad}, demonstrating its potential to explore otherwise inaccessible regions of thermal iDM parameter space\footnote{Complementary studies have explored the sensitivity of the DUNE far detector to MeV-scale dark matter, particularly in scenarios involving boosted dark matter, see e.g. Refs.~\cite{Kim:2016zjx, Berger:2019ttc, DeRoeck:2020ntj, Acevedo:2024wmx, Diurba:2025lky}. In contrast, the scenario considered here focuses on dark matter produced at accelerator-based experiments via a light $U(1)_D$ mediator.}.
 
The rest of the paper is organized as follows. In the next section the model is introduced and its free parameters are identified.  The DM phenomenology relevant for our analysis is also discussed in this section, where the variation of the relic density with the model's parameters is analyzed. Section~\ref{sec:DUNE} provides a detailed description of DM production and detection at DUNE experiment.  Section~\ref{dec:results} presents our main results. There, the viable regions of this model are determined and analyzed, and the detections prospects are investigated. Finally, our conclusions are presented in Section~\ref{sec:conclusions}. 

%%%%%%%%%%%%%%%%%%%%%%%%%%%%%%%%%%%%%%%%%%%%%%%%%%%%%%%%%%%%%%%%%%%%%%%%%
\section{Inelastic DM model}\label{sec:Model}
%%%%%%%%%%%%%%%%%%%%%%%%%%%%%%%%%%%%%%%%%%%%%%%%%%%%%%%%%%%%%%%%%%%%%%%%%
The model extends the SM gauge group through a $U(1)_{D}$ gauge symmetry, where all SM fields are singlets under it. The model also enlarges the particle content with one singlet Dirac fermion, $\psi=\psi_L+\psi_R$,  and one complex scalar field, which in the unitary gauge can be expressed as $\phi = (v_\phi + \hat{h}')/\sqrt{2}$. Their $U(1)_D$  charges  are $1$ and $2$, respectively, in such a way Yukawa interaction terms between them are allowed by the dark symmetry. 
\subsection{Interactions and mass spectrum}
The relevant lagrangian of the model involves several terms, 
\begin{align}
    \mathcal{L}&\supset\mathcal{L}_\psi + \mathcal{L}_{\rm gauge} + \mathcal{L}_{H,\phi}-\mathcal{V}_{H,\phi},
\end{align}
where $\mathcal{L}_\psi$ and $\mathcal{L}_{\rm gauge}$ are the lagrangian associated to the dark fermion and dark ($\hat{X}^\mu$) and hypercharge ($\hat{B}^\mu$) gauge bosons, respectively, while $\mathcal{V}$ denotes the scalar potential of the model. This can be cast as 
\begin{align}\label{eq:V}
    \mathcal{V}_{H,\phi} & = \lambda_{H} \left( H^{\dagger} H - \frac{v_{H}^{2}}{2} \right)^{2} + \lambda_{\phi} \left( \phi^{\ast} \phi - \frac{v_{\phi}^{2}}{2} \right)^{2} + \lambda_{\phi H} \left( H^{\dagger} H - \frac{v_{H}^{2}}{2} \right) \left( \phi^{\ast} \phi - \frac{v_{\phi}^{2}}{2} \right).
\end{align}
Here the SM Higgs doublet is $H = \left(0,(v_{H} + \hat{h}/ \sqrt{2} )\right)^{T}$. After the $U(1)_D$ and electroweak symmetry breaking, the two CP-even scalars mix giving rise to two massive Higgs bosons $(h,h')$, which are related to the interaction fields through the transformation
\begin{equation}
    \begin{pmatrix}
        \hat{h} \\
    \hat{h^{\prime}}
    \end{pmatrix}= 
    \begin{pmatrix}
        \cos\theta & \sin\theta \\
      -\sin\theta & \cos\theta
    \end{pmatrix}
    \begin{pmatrix}
        h \\ 
        h^{\prime}
    \end{pmatrix}.
\end{equation}

The interaction terms involving just the $U(1)$ gauge bosons are~\cite{Bauer:2018onh,Fabbrichesi:2020wbt}
\begin{equation} \label{eq:LG}
    \mathcal{L}_D=\, -\frac14\, \hat{X}_{\mu\nu}\, \hat{X}^{\mu\nu} -\frac14\, \hat{B}_{\mu\nu}\, \hat{B}^{\mu\nu} -\frac14\, \hat{W}_{\mu\nu}\cdot \hat{W}^{\mu\nu}  + \frac{\epsilon}{2}\, \hat{X}_{\mu\nu}\, \hat{B}^{\mu\nu}.
\end{equation}
where $\hat{X}_{\mu\nu} \equiv \partial_\mu \hat{X}_\nu - \partial_\nu \hat{X}_\mu$ and $\hat{B}_{\mu\nu}\equiv \partial_\mu \hat{A}_\nu - \partial_\nu \hat{A}_\mu$.  Here the $\epsilon\ll1$ term allows for a kinetic mixing between $\hat{X}^\mu$ and $\hat{B}^\mu$. Besides those, the kinetic terms for scalars induce the extra terms
\begin{align}
    \mathcal{L}_{H,\phi}&=|D_\mu H|^2+|D_\mu\phi|^2\supset \frac{1}{2}m_{\hat{Z}}^2\hat{Z}_\mu\hat{Z}^\mu(1+\hat{h}/v_H)^2+\frac{1}{2}m_{\hat{X}}^2\hat{X}_\mu\hat{X}^\mu(1+\hat{h'}/v_\phi)^2,
\end{align}
with  $m_{\hat{Z}}=\tfrac{1}{2}\sqrt{g'^2+g^2}\,v_H$ and $m_{\hat{X}}=2\,g_X\,v_\phi$. Focusing on the regime $\epsilon\ll1$ and $\hat{M}_X^2/\hat{M}_Z^2\ll1$, and introducing $\epsilon'=\epsilon\, c_W$, the transformation from the gauge to the mass basis becomes~\cite{Babu:1997st, Frandsen:2011cg}
 \begin{align}
 \begin{pmatrix}
        \hat{B}_\mu\\
        \hat{W}_{3,\mu}\\
        \hat{X}_\mu
    \end{pmatrix}&=
    \begin{pmatrix}
        c_W & -s_W & -c_W\,\epsilon'\\
        s_W & c_W & -s_W\,\epsilon'\\
        0 & t_W\,\epsilon' & 1
    \end{pmatrix}
     \begin{pmatrix}
        A_\mu\\
        Z_{\mu}\\
        A'_\mu
    \end{pmatrix},
\end{align}
or equivalently 
 \begin{align}
 \hat{A}_\mu&=c_W \hat{B}_\mu +s_W \hat{W}_{3,\mu}=A_\mu-\epsilon'A'_\mu,\\
\hat{Z}_\mu&=-s_W \hat{B}_\mu +c_W \hat{W}_{3,\mu}=Z_\mu,\\
\hat{X}_\mu&=A'_\mu+t_W\,\epsilon'Z_\mu. 
\end{align}
The non-zero gauge boson masses at order $\mathcal{O}(\epsilon^2)$ turn to be 
 \begin{align}
     M^2_Z&\approx\hat{M}_Z^2+\mathcal{O}(\epsilon^2),\\
     M^2_{A'}&\approx\hat{M}_X^2+\mathcal{O}(\epsilon^2),
 \end{align}
 
Finally, the dark fermion lagrangian is given by  
\begin{align}
    \mathcal{L}_\psi=&\,i\bar{\psi}\slashed{D}\psi  - m_{D}\bar{\psi}\psi - \frac{1}{\sqrt{2}}\left[f\phi^\dagger\bar{\psi}^{c}\psi + f'\phi^\dagger\bar{\psi}^{c}\gamma_5\psi+\text{h.c.}\right]\\
    =&\,i\bar{\psi}_R\slashed{D}\psi_R +i\bar{\psi}_L\slashed{D}\psi_L - \left[m_{D}\bar{\psi}_R\psi_L+\frac{1}{2}m_L\bar{\psi}_L^{c}\psi_L + \frac{1}{2}m_R\bar{\psi}_R^{c}\psi_R+\text{h.c.}\right]\nonumber\\
    &\,-\frac{1}{2v_\phi}\hat{h}^{\prime}\left[m_L\bar{\psi}_L^{c}\psi_L + m_R\bar{\psi}_R^{c}\psi_R+\text{h.c.}\right],
\end{align}
with $D_{\mu} = \partial _{\mu}  - i g_{X}\hat{X}_{\mu}$, $m_L=(f-f')v_\phi$ and $m_R=(f+f')v_\phi$. In what follows we assume parity conservation in the dark sector such that $f'=0$, that is $m_L=m_R$. 
The spontaneous breaking of the dark symmetry splits the Dirac fermion into two Majorana fermions, 
\begin{equation}
    \chi_{1L}=\frac{1}{\sqrt{2}}(\psi_L-\psi_R^c) = \frac{1}{\sqrt{2}}(\psi-\psi^c)_L,\,\,\,\chi_{2L} = \frac{1}{\sqrt{2}}(\psi_L+\psi_R^c)= \frac{1}{\sqrt{2}}(\psi+\psi^c)_L,
\end{equation}
with masses  
\begin{equation}
    m_{2,1} = m_{D} \pm m_L=m_{D} \pm f v_{\phi}.
\end{equation}

All in all, the model lagrangian in the mass basis comprises the following interaction terms
\begin{align}
\label{eq:equationForChi}
    \mathcal{L}&\,\supset\,\frac{f}{2} (-\sin\theta h+\cos\theta h') \left( \bar{\chi}_{1}\chi_{1} - \bar{\chi}_{2}\chi_{2}\right)+\frac{i}{2}\, g_X (A'_\mu+s_W\,\epsilon Z_\mu) \left( \bar{\chi}_{2} \gamma^{\mu} \chi_{1} - \bar{\chi}_{1} \gamma^{\mu} \chi_{2}\right)\nonumber\\
    &\,+\, \frac{1}{2} m_{A'}^2 A'_\mu A'^\mu\frac{1}{v_\phi}(\sin\theta h-\cos\theta h')\left[\frac{1}{v_\phi}(\sin\theta h-\cos\theta h')-2\right]-\epsilon\, e\, A'_\mu\, J^\mu_{\rm EM},
\end{align}
with $J_{\rm EM}^\mu$ indicating the electromagnetic current. Altogether, this model setup is characterized by seven independent free parameters, which can be chosen as
\begin{align}
    m_1,\,\Delta\equiv m_2-m_1,\,\epsilon,\, \alpha_X\equiv\frac{g_X^2}{4\pi},\,R\equiv \frac{m_{A'}}{m_1},\,m_{h'},\,\sin\theta.
\end{align}
All remaining parameters can be expressed in terms of these through the relations given above. 
For instance, 
\begin{align}\label{eq:non-free-par}
    & v_\phi=\frac{R\,m_{1}}{2g_X},\,\,\,f=\frac{\Delta\,g_X}{R\,m_1},\,\,\,\lambda_{H}= \frac{m_{h}^{2}\cos^2\theta + m_{h'}^2\sin^2\theta}{2v_{H}^{2}},\nonumber\\
    &\lambda_{\phi} = \frac{m_{h}^{2}\sin^2\theta + m_{h^{\prime}}^{2}\cos^2\theta}{2v_{\phi}^{2}},\,\,\, \lambda_{\phi H} = \frac{(m_{h^{\prime}}^{2} - m_{h}^{2})\sin2\theta}{2v_{H}v_{\phi}}.
\end{align}
We impose the following theoretical  constraints on the scalar potential parameters~\cite{Belanger:2014bga}:
\begin{align}
    &\lambda_\phi>0, \,\,\,\lambda_\phi<\pi,\,\,\,|\lambda_{\phi H}|<4\pi,\,\,\,
    \left|6\lambda_H+\lambda_{\phi}\pm \sqrt{(6\lambda_{H}-\lambda_\phi)^2+8\lambda^2_{\phi H}}\right|<16\pi,\nonumber\\
    &\lambda_{\phi H}+2\sqrt{\lambda_H\lambda_\phi}>0,\,\,\,\lambda_{\phi H}v_\phi^2 v_H^2+\lambda_\phi v_\phi^4 >0,\,\,\,\,\,\lambda_H v_H^4+\lambda_{\phi H}v_\phi^2 v_H^2 > 0.
\end{align}
Besides, the Yukawa and gauge couplings are subject to the perturbativity constraint 
\begin{align}
    f<\sqrt{4\pi},\,\alpha_X<1.
\end{align}
Requiring perturbativity leads to constraints on the mass hierarchy of the model. In particular, for the parameter space of interest the dark Higgs $h^\prime$ cannot be significantly heavier than the dark photon $A^\prime$. Concretely, assuming $\epsilon$ and $\theta$ are small with $\alpha_X\sim 0.1$, one obtains $m_{h^\prime}/m_1\,\lesssim\,R$. 
On the experimental side, because of the mixing in the scalar sector, the SM Higgs interactions get modified by a factor of $\cos\theta$ and new interactions of $h^\prime$ with the SM charged fermions are induced. Hence, from direct searches it follows that the condition $|\sin\theta| \lesssim 10^{-4}$ must be fulfilled for $50\,{\rm MeV}<m_{h^\prime}<5\,{\rm GeV}$~\cite{Ferber:2023iso}. In our analysis we focus in the regime where all the new particles lie in the sub-GeV scale, and therefore we fix $\sin\theta=5\times10^{-5}$. 

\subsection{DM relic abundance}
In this model, the DM particle couples diagonally to the scalars $h$ and $h'$. As a consequence, DM annihilation can proceed into SM charged fermions, $\chi_1 \chi_1 \to \bar{f} f$, as well as into dark scalars, $\chi_1 \chi_1 \to h' h'$ ($\chi_2 \chi_2 \to h' h'$ may be also possible). Although the latter channel is $p$-wave suppressed, it dominates in the region of parameter space of interest, since the former—while not velocity suppressed—is strongly suppressed by both the small SM Yukawa couplings and the scalar mixing angle. This hierarchy of annihilation channels is crucial for evading the stringent CMB constraints on the DM annihilation cross section.

In addition, within the parameter regions considered here, the formation of bound states of $\chi_1$ or $\chi_2$ via radiative emission of $h'$~\cite{Ko:2019wxq} is kinematically forbidden for non-relativistic $\chi_1$, that is, as long as $f^4m_1/(64\pi^2)<m_{h'}$. 

In contrast, the interaction mediated by the dark photon is non-diagonal, giving rise to the $s$-wave process $\chi_1 \chi_2\to\bar f f$, suppressed by $\epsilon^2$ and at late times by the number density of $\chi_2$ which is controlled by the mass splitting $\Delta$. The other possible $s$-wave channel, $\chi_1 \chi_1\to A'A'$ is kinematically closed in our analysis, since $R\geq3$ and $\Delta<m_1$. These features play a key role when evading the CMB constraints. 

As a result, the DM relic abundance is mainly determined by the interplay of the standard inelastic process $\chi_1 \chi_2\to\bar f f$ and the new channel involving the dark Higgs, $\chi_1 \chi_1 \to h' h'$. By considering small values of $\epsilon$ or increasing $R$, the annihilation rate into SM fermions --which is always kinematically open-- can be significantly reduced. At the same time, annihilation into dark Higgs pairs is further suppressed in the forbidden annihilation regime $m_1<m_{h^\prime}\lesssim1.5\,m_1$~\cite{DAgnolo:2015ujb}. 

\begin{figure}[t]
    \centering
    \includegraphics[width=0.45\linewidth]{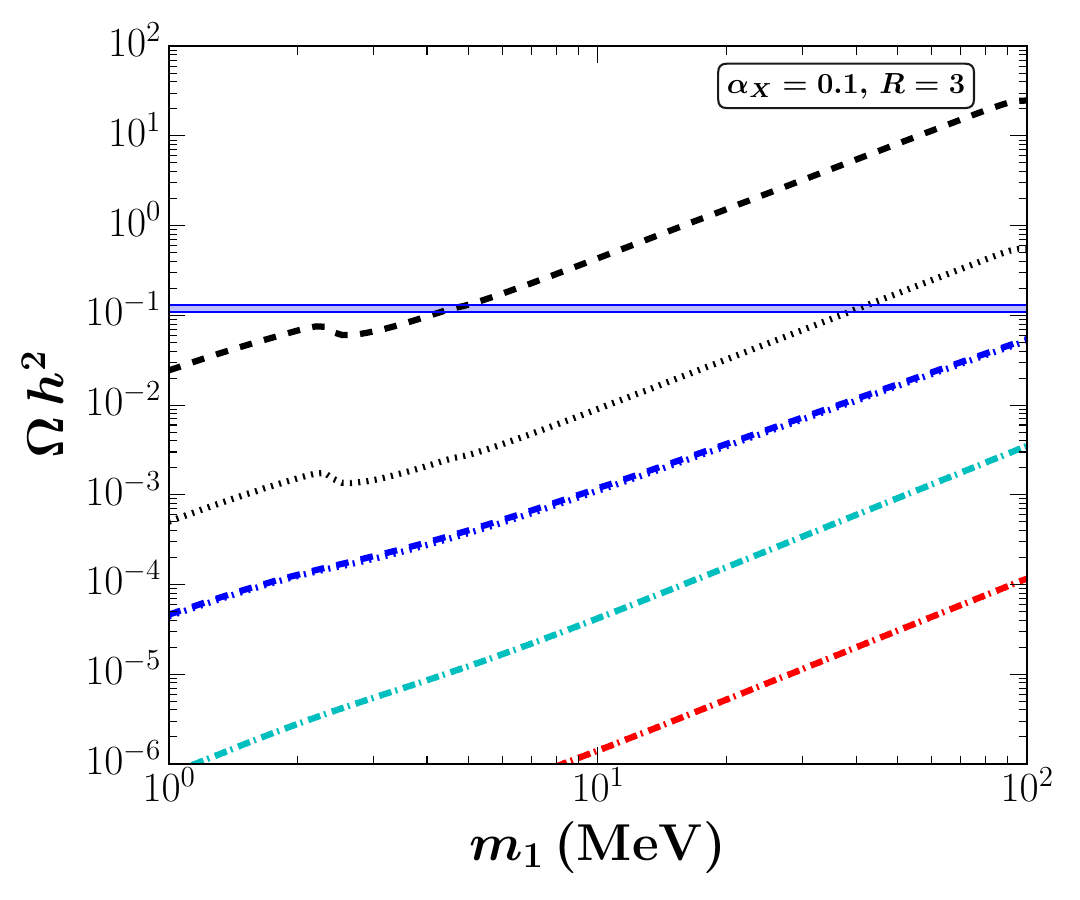}
    \includegraphics[width=0.45\linewidth]{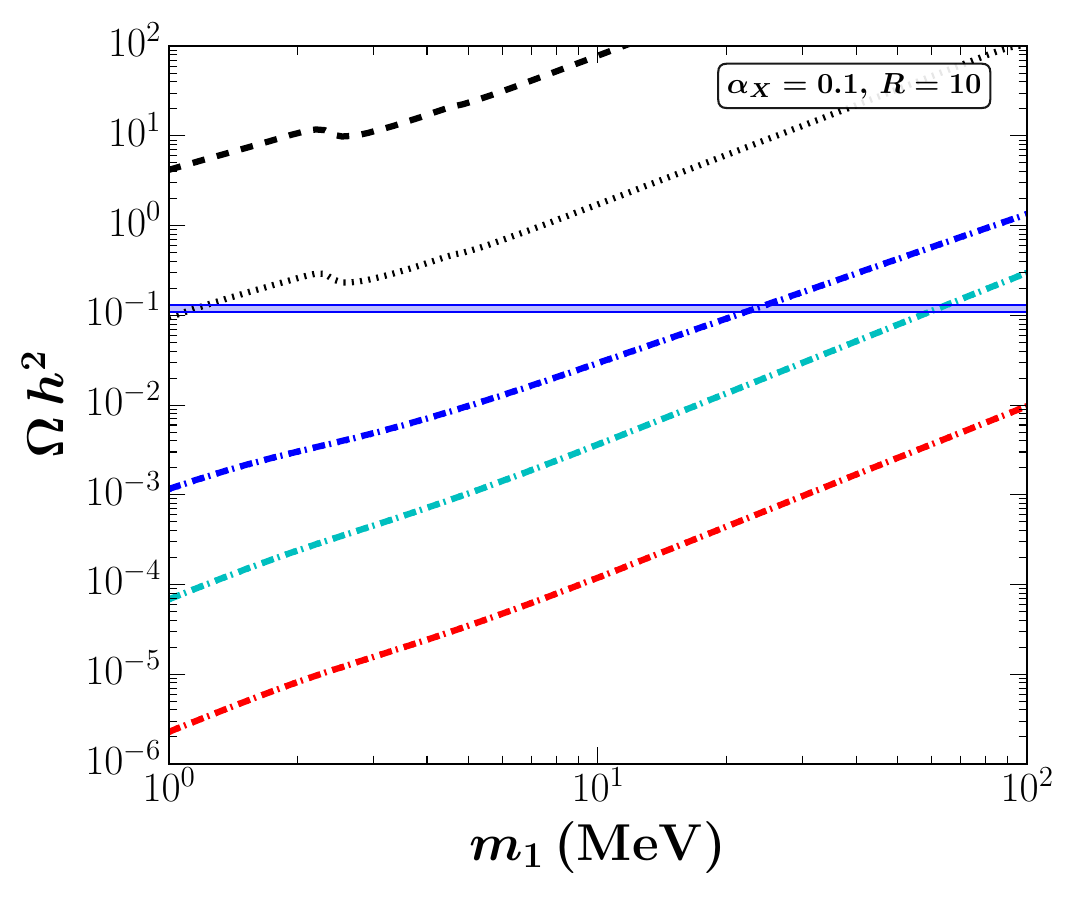}\\
    \includegraphics[width=0.45\linewidth]{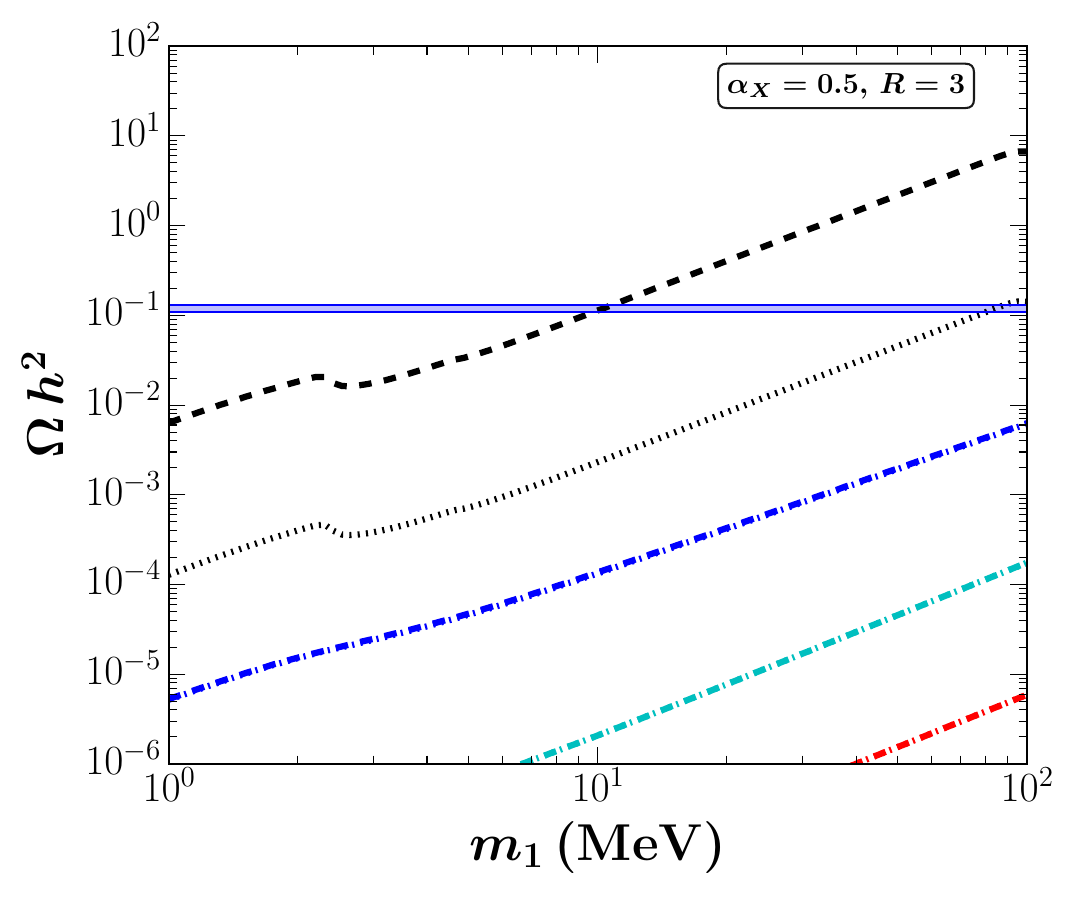}
    \includegraphics[width=0.45\linewidth]{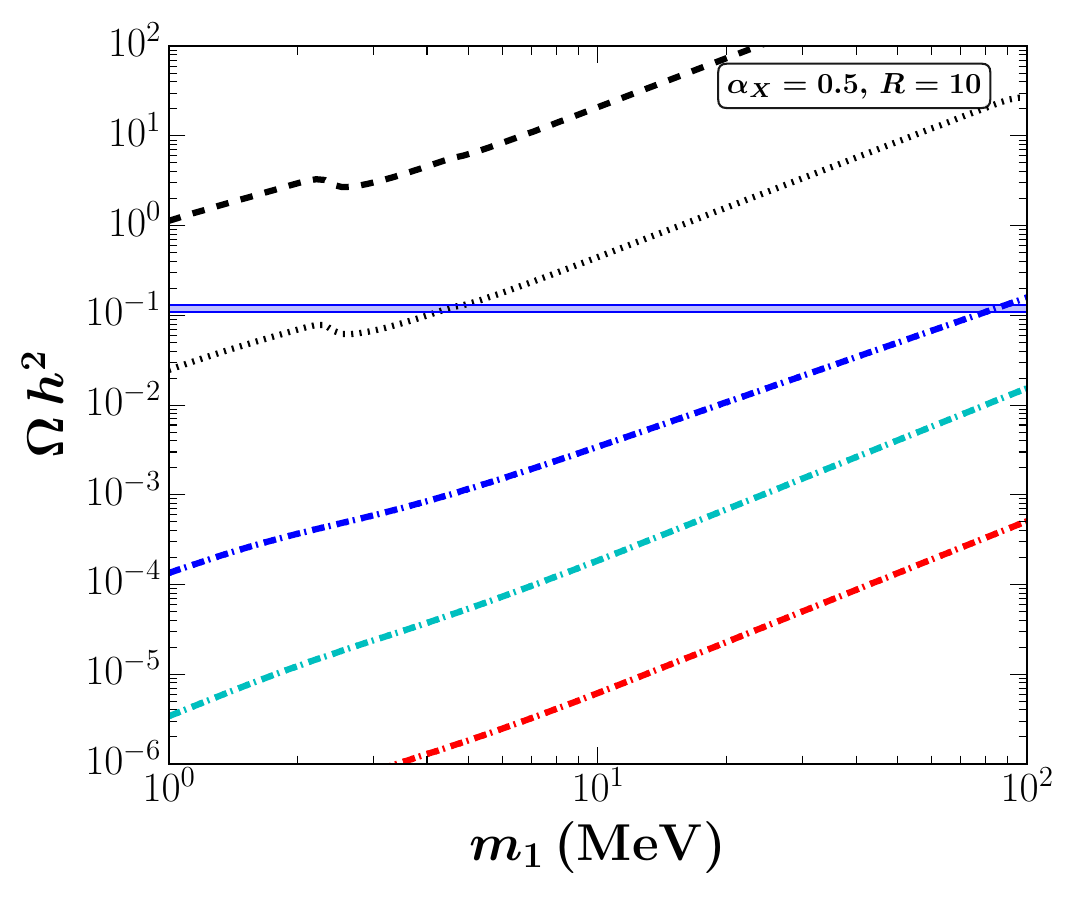}   
    \caption{DM abundance as a function of its mass. The cyan, red, blue and black lines correspond to $m_{h^\prime}/m_1=0.01,1,1.2,2$, respectively, while dashed (dotted) lines denote $\epsilon=10^{-5}\,(10^{-4})$. The left (right) panels are obtained for $R=3\,(10)$ whereas the top (bottom) panels correspond to $\alpha_X=0.1\,(0.5)$. The remaining free parameters are fixed to $\Delta=0.1\, m_1$ and $\sin\theta=5\times10^{-5}$.}
    \label{fig:omega-vs-m1}
\end{figure}

To illustrate how the different parameters affect the relic abundance, we vary the free parameters of the model as shown in Fig.~\ref{fig:omega-vs-m1}.  
The DM relic abundance is computed using {\tt micrOMEGAs}~\cite{Belanger:2018ccd}, with the model implemented via the {\tt LanHEP}~\cite{Semenov:2014rea} and {\tt FeynRules}~\cite{Alloul:2013bka} packages. We emphasize that {\tt micrOMEGAs} numerically solves the Boltzmann equation for $\chi$, automatically accounting for all relevant processes contributing to the relic abundance, including forbidden annihilation channels. 

We first analyze the impact of varying $\epsilon$ and the mass ratio $m_{h\prime}/m_1$ for $\alpha_X=0.1$ and $R=3$ (left top panel).  The cyan, red, blue and black lines correspond to $m_{h^\prime}/m_1=0.01,1,1.2, 2$, respectively, while dashed (dotted) lines denote $\epsilon=10^{-5}\,(10^{-4})$. 
We find that $\chi$ is over-abundant in almost the whole mass range for $\epsilon=10^{-5}$ and $m_{h^\prime}=2m_1$ (black dashed line), that is, when annihilation into fermions is slightly efficient and the annihilation into dark Higgses is kinematically closed. Reducing the mass ratio to $m_{h^\prime}/m_1=1.2$ (blue dashed line) leads to a decrease of almost three orders of magnitude in the relic abundance, signaling the onset of the forbidden annihilations proceeding from the thermal tail. A further reduction of approximately two orders of magnitude in $\Omega$ is obtained when the mass ratio is decreased to the unity (red dashed line). 
Finally, when annihilation into dark Higgs bosons is kinematically open at all temperatures ($m_{h'}/m_1=0.01$, cyan dashed line), the relic abundance increases again. In this regime, the process $\chi_1 \chi_1 \to h' h'$ is no longer enhanced by thermal effects and proceeds as a purely $p$-wave–suppressed channel. As a result, the annihilation rate at freeze-out is reduced compared to the near-threshold (forbidden) regime, leading to a less efficient depletion of the DM density and hence a larger relic abundance.
Increasing $\epsilon$ to $10^{-4}$ (dotted lines) results in a qualitatively similar behavior. However, a significant additional suppression of $\Omega$ is observed only for $m_{h^\prime}=2m_1$, while no further suppression is obtained for the other mass ratios.   

The effect of a heavier dark photon, corresponding to $m_{A^\prime}=10m_1$, is shown in the top right panel. Since increasing $R$ by a factor of 10/3 reduces the Yukawa coupling $f$ by the same factor (see Eq.~\eqref{eq:non-free-par}), the annihilation rate associated into dark Higgses is correspondingly suppressed. As a result, the relic abundance increases, as can be seen by comparing the cyan, red and blue curves with their counterparts in $R=3$ case. Moreover, the increase in $R$ also suppresses the rate for the $\chi_1 \chi_2\to\bar f f$ process, leading to further enhancement of $\Omega$, as is illustrated by the black curves in both top panels. 

Finally, the impact of increasing the gauge coupling $\alpha_X$ is shown in the bottom panels. In this case, the enhanced annihilation rates lead to a systematic decrease in the relic abundance compared to the corresponding scenarios shown in the top panels.

In obtaining the above results, we assume that $\chi_2$ is cosmologically unstable. This is ensured by the $U(1)_D$ charge assignment, which guarantees the absolute stability of $\chi_1$ while allowing $\chi_2$ to decay. In the parameter region of interest%
\footnote{The two-body decays $\chi_2 \to \chi_1 + X$, with $X = A', Z, h', h$, are kinematically forbidden.}
$\chi_2$ decays dominantly through the three-body channel $\chi_2 \to \chi_1 \bar{f} f$, where $f$ can be electrons, muons, or light quarks, depending on the available phase space. The corresponding lifetime is such that $\chi_2$ does not contribute to the present-day DM abundance.

For the benchmark value $\Delta = 0.1\, m_1$, the minimum DM mass that allows for these decays is $m_1 \simeq 10~\mathrm{MeV}$, resulting in lifetimes shorter than the onset of Big Bang nucleosynthesis. When the decay into fermions becomes kinematically forbidden, $\chi_2$ instead decays radiatively into $\chi_1$ plus three photons. In this case, the corresponding lifetime is of the order of, or shorter than, the time of cosmic microwave background formation for DM masses slightly below $10~\mathrm{GeV}$.

%%%%%%%%%%%%%%%%%%%%%%%%%%%%%%%%%%%%%%%%%%%%%%%%%%%%%%%%%%%%%%%%%%%%%%%%%
\section{Sub-GeV DM at fixed-target experiments}\label{sec:DUNE}
%%%%%%%%%%%%%%%%%%%%%%%%%%%%%%%%%%%%%%%%%%%%%%%%%%%%%%%%%%%%%%%%%%%%%%%%%
In this section, we explore the sensitivity of fixed-target experiments to the parameter space of the model. For concreteness, we focus on the upcoming DUNE experiment. 

DUNE is a long baseline neutrino experiment that uses noble liquid argon time projection chambers (LAr TPC) for neutrino detection~\cite{DUNE:2021tad}. At DUNE, neutrinos are produced by the collision of protons accelerated at energies of 60 GeV or 120 GeV with a carbon target. The experiment will have the most intense neutrino beam with $1.1\times 10^{21}$ protons on target (POT) per year. After the collision, mesons such as pions are produced, which subsequently decay to neutrinos or antineutrinos. The experiment will have several TPCs to detect neutrino interactions. 
Two of them, the far detectors, will be located 1300 km from the collision point, and one, called the LAr cube will be located 574 m from the collision point. The latter is part of the so-called near detector (ND) complex. The LAr cube will have dimensions of 7m$\times$3m$\times$5m with a 67-ton fiducial mass~\cite{DUNE:2021tad}. This complex is fundamental for the characterization of the neutrino beam produced before oscillation.

Although DUNE's main goals relate to neutrino physics, the experiment can be an important tool for probing these sub-GeV dark sectors. The following provides a detailed description of DM production and detection at DUNE.

\subsection{DM production}
DM generation at the experiment relies on the production of $A^{\prime}$, there are mainly three mechanisms:
\begin{itemize}
    \item Drell-Yan: via the process $p \ p \rightarrow A^\prime \rightarrow \chi_1  \ \chi_2 $, this process becomes important for DM masses greater than 1 GeV~\cite{DeRomeri:2019kic, SHiP:2020vbd}, which are larger than the masses considered in this work.
    
    \item Meson decay: Several mesons are produced at the collision point. At DUNE, the charged mesons decay will lead to neutrino production, while the neutral mesons could lead to sub-GeV DM production. The latter arises because neutral scalar mesons such as $\pi^{0}$ and $\eta$ have large branching fractions to photon-photon. One of these could kinetically mix with the dark photon. This dark sector particle is unstable, with its decay to DM being the dominant channel whenever it is available. This arises from the fact that the kinetic mixing $\epsilon$ does not suppress the DM interaction with the dark photon. To ensure that this important channel is open, we will consider the widely studied benchmark scenario $R \geq 3 $, while the mass splitting  $\Delta$ will be kept constant at $0.1m_1$.

\item Proton Bremsstrahlung: The process $p \ N \rightarrow p \ N \ A^\prime$  is another source of dark photon production; again, the production of this gauge boson is followed by a decay into the dark sector particle. Refs.~\cite{SHiP:2020noy,Breitbach:2021gvv,DeRomeri:2019kic} found that this production channel is mostly relevant for dark photon masses $0.5 \lesssim m_{A^\prime}/{\rm GeV} \lesssim 1.0 $, that is, for masses above the neutral scalar meson mass $\eta$.
\end{itemize}

Regarding the relative importance of each production channel, meson decay dominates whenever the corresponding neutral-meson decay is kinematically open. Thus, for $m_{A'}<m_{\pi^0}$, both $\pi^0$ and $\eta$ decays contribute, whereas for $m_{\pi^0}<m_{A'}<m_{\eta}$ the production is dominated by $\eta$ decays. For $m_{A'}>m_{\eta}$, the meson decay channels are closed, and proton Bremsstrahlung becomes the relevant production mechanism. Nevertheless, as will be shown below, proton Bremsstrahlung is not the channel driving the new parameter space that can be explored by DUNE, and it is included here for completeness.

Additional channels to the ones considered above do exist and can be relevant. In DUNE, due to the beam energies, as well as the target material and geometry, there will be large electromagnetic activity, which opens new dark photon production channels. For instance, $e^{\pm}$ bremsstrahlung, $e^{+}e^{-}$ resonant annihilation, associated production $e^{+}e^{-}\rightarrow A' \gamma$, and dark Compton-like processes have been found to yield important contributions to the dark sector flux ~\cite{Celentano:2020vtu,Capozzi:2021nmp,Blinov:2024pza}. In particular, Ref.~\cite{Blinov:2024pza} found that, for $m_{A'}\lesssim 30$ MeV, resonant production can give a larger yield than meson decay. These channels can therefore improve DUNE's signals in the low dark photon mass regime. In this work, we do not include these electromagnetic production channels, since they are not implemented in \texttt{MadDump}~\cite{Buonocore:2018xjk}. Therefore, the DUNE sensitivities we report here should be interpreted as conservative estimates based on the hadronic channels, especially in the low $m_{A'}$ regime.

\subsection{DM detection}
For inelastic DM, there is more than one possible detection signal, depending on the parameter space of interest; this is because the heavier dark sector particle could decay on flight, at the detector, or be stable long enough to scatter with the detector material. The main parameter (though not the only one) to define the different regions is the mass splitting $\Delta$. If $\Delta \geq 2 m_{e}$ with $m_{e}$ the electron mass, the decay channel $\chi_2 \rightarrow \chi_1  \ e^{+} \ e^{-}$ opens and is mediated by an off-shell $A^\prime$. The width decay can be approximated as~\cite{Tsai:2019buq,CarrilloGonzalez:2021lxm,Fitzpatrick:2021cij}
\begin{align}
\label{eq:width_chi2}
  \Gamma_{\chi_{2}} \approx \frac{4 \alpha \alpha_X \epsilon^2 \Delta^5}{15 \pi m_{A^\prime}^4}.
\end{align}
For the stability or decay of $\chi_2$, we can identify three regimes on the $\Delta$ parameter:
\begin{itemize}
\item Large $\Delta$. 
In this case, the heavier dark sector particle is very likely to decay on flight before reaching the ND complex or at the ND complex. If the decay occurs before reaching the TPC, the $\chi_1$ decay product could travel to the LAr and scatter there. If the decay happens at the TPC, the signal to be considered is a separated $e^{+}$ and $e^{-}$. 
The latter is a well-studied signal in many fixed-target experiments and has been used to place restrictions on large mass splittings $\Delta$. In fact, the strongest constraints arise from these decay searches or from re-casting.

\item Intermediate $\Delta$.
With the decay channel to leptons opened, $\chi_2$ should be unstable; however, at DUNE energies, the boost $\beta \gamma$ can dilate its lifetime, allowing it to travel distances longer than those of the ND complex before decaying. If this is the case, for DUNE purposes, $\chi_2$ can be considered stable. This scenario is very similar to the one with a small $\Delta$. In such a case, both $\chi_1$ and $\chi_2$ produced at the collision can travel to the LAr Cube and scatter with the detector material. This scenario can lead to good sensitivity at DUNE since both particles are nearly equally likely to interact at the ND, thus enhancing the statistics. Moreover, this small splitting is currently not as constrained as the large one. 

\item Small $\Delta$. For a mass splitting smaller than $2 m_e$, $\chi_2$ is stable because its only allowed decays are $\chi_2 \rightarrow \gamma \ \gamma \ \gamma \ \chi_1$ and $\chi_2 \rightarrow \nu \ \nu \ \chi_1$~\cite{CarrilloGonzalez:2021lxm}\footnote{Note that the decay $\chi_2\to\chi_1\,\gamma$ is absent at tree level because the Majorana states couple off-diagonally only to the dark photon and the $Z$ boson. As a result, no gauge-invariant dipole operator involving the SM photon is generated at tree level.}. These decays are very suppressed, rendering $\chi_2$ absolutely stable for DUNE's purposes. The signal is the same as the intermediate mass splitting. However, we will not take into account this scenario because it leads to tensions with the predictions of Big Bang nucleosynthesis.  
\end{itemize}
We emphasize that there is not one value of $\Delta$ that separates the difference between intermediate and large splitting regimes. The relevant observable is the decay length of the heavier partner. This, in turn, has a non-trivial dependence on the mass ratio $R$, $\epsilon^2$, and $\alpha_X$. For instance, for $\alpha_X=0.1$, $\Delta=0.1 m_1$, and $R=3.0$, $\chi_2$ behaves as a stable particle even for $\epsilon^2$ as high as $\sim 10^{-4}$. On the other hand, for $\alpha_X=0.5$, $\Delta=0.4  m_1$, and $R=10.0$, $\chi_2$ behaves as a stable particle for $\epsilon^2 \lesssim 10^{-6}$. In this sense, we define the intermediate-splitting regime as the region in which $\chi_2$ is unstable in principle, since the decay channel into charged leptons is open, but is sufficiently long-lived to reach the DUNE ND and scatter before decaying.
This distinction also clarifies the complementarity with visible-decay searches. Visible decay signatures, such as $\chi_2\to\chi_1 e^+e^-$, are particularly interesting because they can yield stronger sensitivity due to their cleaner topology and smaller backgrounds. In fact, several fixed-target experiments have searched for such signals, leading to strong existing constraints in the region of parameter space where $\chi_2$ decays before or inside the detector ~\cite{Mongillo:2023hbs,Izaguirre:2017bqb}. The scattering signal considered in this work is therefore complementary to these searches. It is relevant because it can probe regions where $\chi_2$ is long-lived on DUNE ND scales, which are less constrained.

The intermediate mass-splitting regime offers a statistical advantage for a DUNE analysis since both $\chi_1$ and $\chi_2$ can imprint signals in the LAr cube, and the experiment can probe portions of the parameter space currently unexplored. We therefore focus on this regime. 
The scattering can occur with nucleons or electrons; however, scattering with nucleons in the ND-LAr has been shown to have large backgrounds due to neutrinos~\cite{DeRomeri:2019kic}, and the DM signal cannot compete with them\footnote{Additional detector capabilities may provide enhanced discrimination between neutrino-induced backgrounds and massive particles~\cite{MicroBooNE:2023ldj,MicroBooNE:2023gmv}.}. On the other hand, several works have shown that DUNE can have sensitivity to electron scattering despite the large neutrino background~\cite{DeRomeri:2019kic,Breitbach:2021gvv,Celentano:2020vtu}. This technique utilizes the fine angular and energy resolution of the LAr technology. For a dark sector particle to scatter with electrons, this model has two possible channels (see Fig.~\ref{fig:scattering}):

\begin{figure}[!h]
    \centering
        \includegraphics[width=0.27\linewidth]{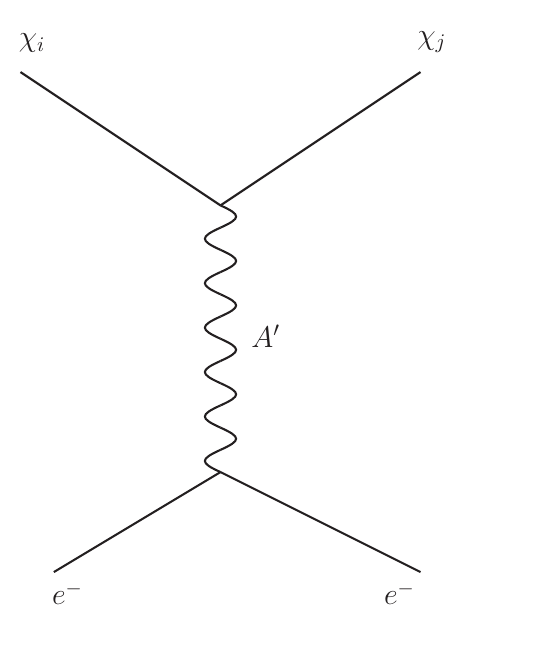}\hspace{1.5cm}
    \includegraphics[width=0.27\linewidth]{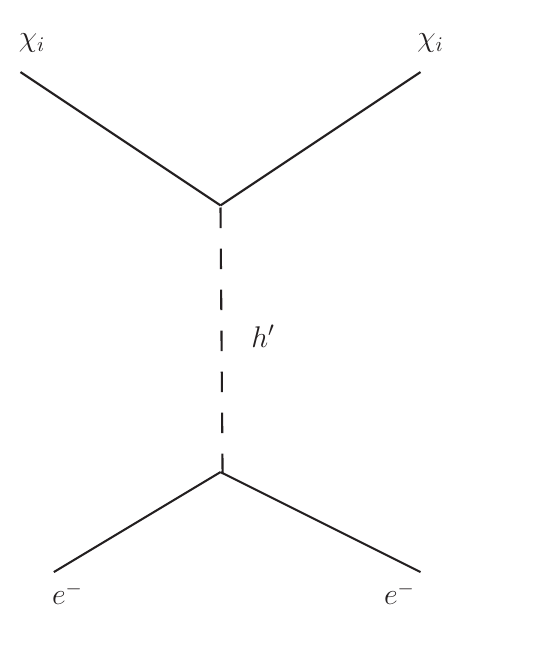}\\
    \caption{\centering Feynman diagrams showing the inelastic scattering channel for DUNE detection (left panel) and the elastic scattering channel (right panel). For the left panel, if $i=1(2), j=2(1)$, there is an up(down)-scattering with electrons. For the right panel, the elastic scattering can occur with $i=1$ or 2.}
    \label{fig:scattering}
\end{figure}

\begin{itemize}
    \item Dark photon mediated:
    This channel is enabled by the off-diagonal coupling of the dark sector particles with $A^\prime$ and the interaction of this with electromagnetic current, as a result, a dark sector particle scatters with an electron via up-scattering $\chi_1 \ e^{-} \rightarrow \chi_2 \ e^{-}$ or down-scattering $\chi_2 \ e^{-} \rightarrow \chi_1 \ e^{-}$. Since we keep the mass splitting $\Delta= 0.1m_1$, we do not expect considerable differences between the kinematics and event rates of these two channels. As such, both event rates are proportional to $\epsilon^2 \alpha_{X}$. 
    \item Dark Higgs mediated:
    This channel is enabled by the diagonal coupling of the dark sector particles to $h^{\prime}$ and the mixing of $\hat h^{\prime}-\hat h$, which renders the event rates proportional to $\alpha_{f} y_{e}^2 \ \theta^2 $~\cite{Duerr:2020muu}, where $y_{e}$ is the electron Yukawa coupling. 
\end{itemize}

\subsection{Simulation of DM interactions at the ND}
To probe the ND's LAr capabilities for exploring this model, we consider an Ar cube of 67-ton fiducial mass. 
To simulate DM production and detection, we first employ \texttt{Pythia 8.3}~\cite{Bierlich:2022pfr} to model the proton-target collision and the subsequent production of neutral mesons ($\pi^0$,$\eta$), assuming a beam energy of 120 GeV, we use the flag \texttt{SoftQCD:all} which is appropriate for the energies expected at DUNE. A seven-year DUNE run will yield $7.7\times 10^{21}$ protons on target (POT), an the \texttt{Pythia} simulation yields an expected meson production of $4.1$ $\pi^0$ and $0.5$ $\eta$ per POT, in agreement with Ref.~\cite{DeRomeri:2019kic}. Additionally, we compared our meson kinematic distributions with those obtained for DUNE in Ref.~\cite{Celentano:2020vtu}, and we find a good agreement with our results. We used the meson kinematics results from \texttt{Pythia 8.3} to simulate their decay to DM. This is done by \texttt{MadGraph}~\cite{Alwall:2014hca} and its plugin \texttt{MadDump}~\cite{Buonocore:2018xjk}; the result is a Monte Carlo simulation of the DM flux at the LAr detector and the DM interactions with it. Moreover, \texttt{MadDump} provides all the relevant kinematics information for the DM produced and the recoiled electron. It is important to add that \texttt{MadDump}~\cite{Buonocore:2018xjk} also has the capability of simulating dark photon production via proton bremsstrahlung. We included this production channel, ensuring that 
\begin{align}
E_{p},E_{A'},E_{p}-E_{A'} \gg m_p,m_{A^\prime},|p_T|,
\end{align}
where $E_{p}$ and $E_{A^\prime}$ are the incoming proton energies and dark photon energies, respectively. While $|p_T|$ is the outgoing proton momentum in the transverse direction to the beam. These restrictions are important to maintain the validity of the Fermi-Williams-Weizsacker approximation~\cite{Blumlein:2013cua}, which is the one implemented in \texttt{MadDump}~\cite{Alwall:2014hca}.

Regarding DM detection, we considered the two possible mediators for this interaction $h^{\prime}, A^{\prime}$ as explained above; nevertheless, we found that the cross-section for events mediated by a dark Higgs was very suppressed compared to the dark photon this is because the square of the electron Yukawa coupling is of the order of $ \sim 10^{-12}$ while even for the largest allowed values of the mixing angle $ \theta^2 \sim 10^{-8}$. On the other hand, the couplings that enter in the scattering mediated by $A^{'}$ are suppressed by $\epsilon^2 $, with the smallest values considered in this work being of the order $10^{-11}$. This puts the scattering through $h^{\prime}$ several orders of magnitude below the one mediated by $A^{\prime}$. Subsequently, all our analyses will be performed only with the mediator $A^{'}$

For the simulation, we will focus on the following parameters: $\alpha_X=0.5,0.1$, $m_{A^{'}}=R \ m_1$, where $3 \leq R \leq 35$.  Many works studying dark sector models consider only the case of $R=3$; which ensures that the dark photon decays on-shell. However, there is no theoretical reason to ignore other values of $R$. Works such as Refs.~\cite{Fitzpatrick:2020vba,Fitzpatrick:2021cij} have considered the case of $R\lesssim 2$, while in Ref.~\cite{Bjorken:2009mm} it was shown the importance of departing from that one value of $R$, and the great capability of fixed-target experiments to explore larger ratios. On the other hand, for the mass splitting we focus on the case $\Delta=0.1m_1$ while maintaining $ \Delta \gtrsim 2 m_{e} $.

With the Monte Carlo results, we first calculate the stability of $\chi_2$. To do this, for every value of $m_{A^\prime}$ we calculate the boost $\gamma$ and velocity $\beta c$ of every $\chi_2$ produced, which yields a distribution. Now, the probability that the unstable particle will decay at a given distance $d$ is 
\begin{align}
P(d)=1-e^{-d/(c\beta \gamma \tau)},    
\end{align}
where $\tau$ is the particle's lifetime calculated in Eq.~\eqref{eq:width_chi2}, and we use $d=581$ m, the distance from the collision to the end of the argon cube. For each particle we calculate $P(581 \ {\rm m})$, and we consider that particles with a $P(581 \ {\rm m}) \leq 10\%$ behave as stable particles. With each $\chi_2$'s probability at hand, we then calculate for every value of the parameter space ($\Delta$, $\epsilon^2$, $m_{A^\prime}$) what percentage of $\chi_2$'s behave as stable. Suppose the percentage is larger than $85\%$, in that case, we consider that such a parameter space point yields the same sensitivity as a completely stable $\chi_2$. This value is conservative because we have checked that going as low as $70\%$ does not greatly affect our results. Thus, we can define regions in the $\epsilon^2-m_1$ plane for different values of where an unstable $\chi_2$ yields the same sensitivity as a fully stable particle. In fact, we find that for the interesting ranges of the parameters, $\chi_2$ behaves as stable in the distances relevant for the ND complex. Hence, we do not include any further discussions on the stability of the heavier partner.

To probe DUNE's sensitivity, we considered the on-axis configuration running for 7 years. Additionally, we consider a configuration proposed in Ref.~\cite{Brdar:2022vum}, where protons are directed onto a beam dump instead of the carbon target. This reduces the neutrino flux by three orders of magnitude, thereby enabling sensitivity to BSM models with only a few months of data.

 Using the output of \texttt{MadDump}, we obtain, for each event with a scattered electron, its energy $E_e$ and scattering angle $\theta$ with respect to the beam axis. These kinematic variables are used to characterize the DM signal.

In the presence of large backgrounds, several SM processes can mimic the electron recoil induced by DM interactions. Following the convention adopted in previous studies~\cite{Mathur:2021trm}, we treat elastic neutrino–electron scattering ($\nu$–$e$) as the irreducible SM background, while charged-current quasi-elastic neutrino scattering (CCQE) and events with misidentified $\pi^0$ are collectively referred to as background (BKG) processes.

The kinematic variable $E_{e} \theta^2$ provides an excellent handle to study scattering with electrons because the kinematics of the interaction dictate that:
 \begin{align}
     1-\cos(\theta_e) \simeq \frac{m_e}{E_e}(1-y),
 \end{align}
 where $y$ is the inelasticity parameter~\cite{Chakraborty:2024xxc}. DM particles and neutrinos will have a large boost along the beam direction, which leads to a natural cut on the variable $E_{e} \theta^2 \lesssim 2 m_{e}$, and thus it becomes an excellent handle to reject CCQE~\cite{MINERvA:2015nqi,Marshall:2019vdy, Chakraborty:2024xxc,DeRomeri:2019kic}. 
To ensure a realistic analysis, detector uncertainties must be taken into account; in particular, finite energy and angular resolution play a crucial role. In this regard, we apply a Gaussian energy smearing with $\sigma_{E}=0.1$. The ND-LAr detector is expected to have an angular resolution of $\sigma_{\theta}=1^{\circ}$. Angular smearing is more complex than energy smearing, since a deviation in the measured angle from the true angle affects both the polar and azimuthal components – we follow the procedure outlined in Ref.~\cite{Mathur:2021trm} to incorporate this effect. With the simulated energy and angle, including smearing, we calculate the $E_{e} \theta^2$ for each event. We only consider events where $0.5 \times 10^{-3}\,{\rm MeV\, rad}^2 \leq E_{e} \theta^2 \leq 3.0\,{\rm MeV\,rad}^2$.
Since the DM signal, if present, occurs within large backgrounds, we consider the case where no DM signal is present, and the case where DM signals are mixed with background. Although it has been considered that by imposing all $E_{e}\theta^2\leq2m_{e}$, large portions of the CCQE background can be rejected, this no longer applies perfectly when detector uncertainties are considered due to the angular smearing. For all possible backgrounds,  in Ref.~\cite{Mathur:2021trm}, considering that the experiment runs equal amounts of time in neutrino and antineutrino mode.  

\subsection{Statistical analysis for DUNE's sensitivity}

\begin{figure}[h]
    \centering
    \includegraphics[scale=0.4]{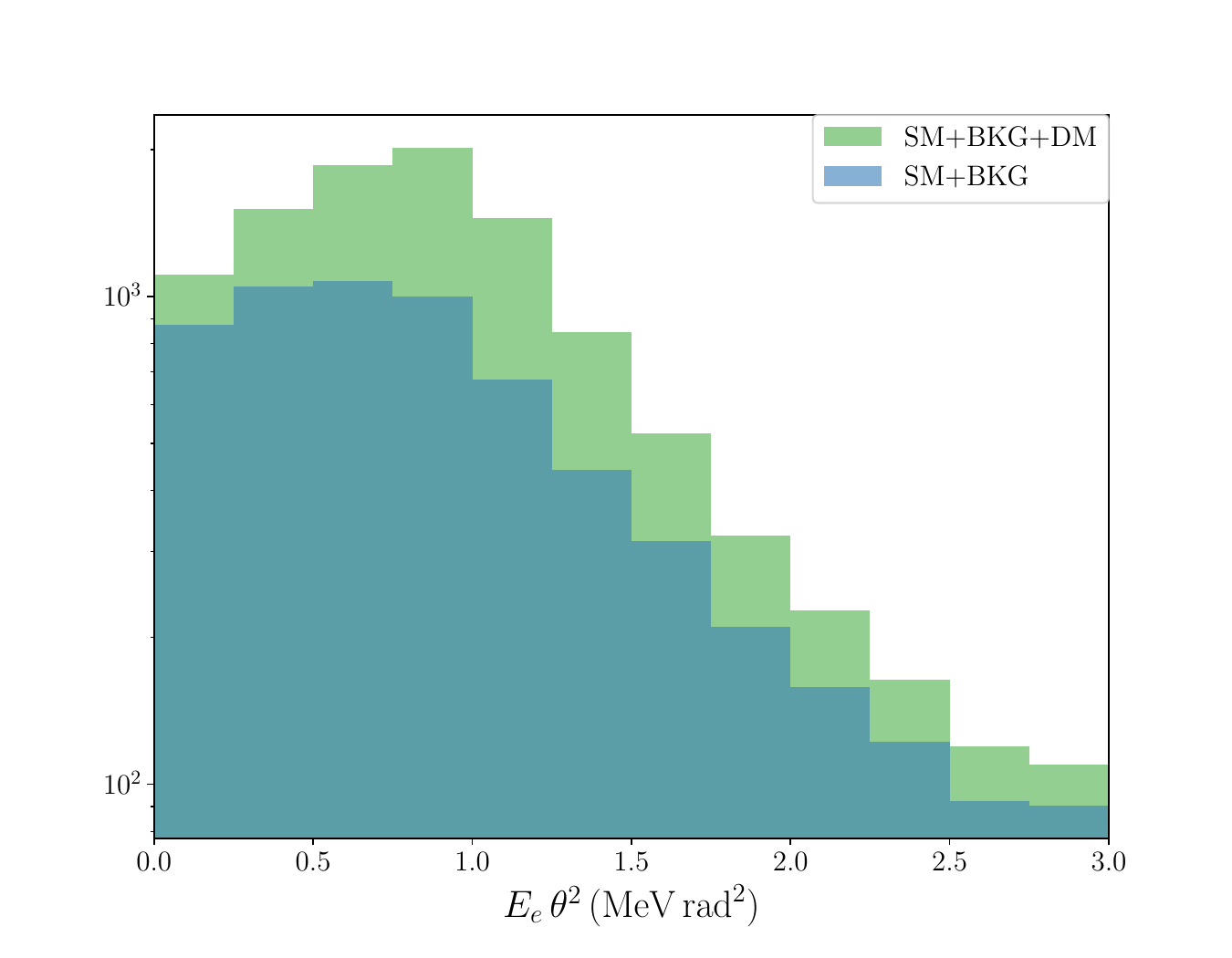}
    \caption{Histogram showing expected number of events for different $E_{e} \theta^2$ for the case of SM + BKG  only, and SM + BKG + DM. The benchmark point considered is $m_{A^\prime}=0.05$ GeV, $\epsilon^2=10^{-8}$, while $\alpha_{X}=0.5$ }
    \label{fig:SM-DM-histo}
\end{figure}

With the scattered electrons $E_{e} \theta^2$ in hand for $\Delta=0.1 m_1$, $0.01\leq m_{A^\prime}/{\rm GeV} \leq 1.0$, and $3 \leq R \leq 35$, we obtain histograms for SM only and SM+DM with 12 bins of data. A sample of such histogram is shown in Fig.~\ref{fig:SM-DM-histo}, where the effect of smearing is tangible. For the analysis, we have considered that for the on-axis setting, the detector runs 3.5 years in neutrino mode and 3.5 years in anti-neutrino mode. 
For the beam-dump configuration, we assume that the total number of background events capable of mimicking the DM signal (SM + BKG) is reduced by three orders of magnitude relative to that in the on-axis configuration. We further assume an on-axis running time of six months.

To explore DUNE's sensitivity to $\epsilon^2-m_1$, we perform a $\chi^2$ test with pull parameters. The pull parameters are introduced to account for the statistical uncertainties in the experiment. A normalization uncertainty of $5\%$ related to the flux, a detector uncertainty, and a background (CCQE+miss identified pions) uncertainty of $10\%$. This scenario was considered in Ref.~\cite{Ballett:2019xoj}, and we follow their notation closely. Taking this into account, our $\chi^2$ test is:
\begin{align}
    \chi^2=\sum_{j}\sum_{i}\left(\frac{(N_{obs_{ij}}-(1+\alpha)N_{{\rm SM}_{ij}}-(\alpha+\beta)N_{{\rm BKG}_{ij}})^2}{N_{{\rm obs}_{ij}}} \right)+ \left( \frac{\alpha}{\sigma_{\alpha}} \right)^2 + \left(\frac{\beta}{\sigma_{\beta}} \right)^2,
\end{align}
where $N_{\rm SM}$ refers to number of $\nu-e$ scattering events, $N_{\rm BKG}$ refers to CCQE and miss identified pions, and $N_{\rm obs}=N_{\rm DM}+N_{\rm SM}+N_{\rm BKG}$ to the number of DM-e scattering events. The index $i$ runs over the 12 histogram bins, the index $j$ runs over $\nu$ and $\bar{\nu}$. 
Next, we perform a $\chi^2$ minimization to calculate the pull parameters, and use the $\chi^2$ to obtain the p-value for a given $\epsilon^2-m_{A^{\prime}}$. We look for $\epsilon^2$ where the experiment can exclude the model at $90\%$ confidence level.

%%%%%%%%%%%%%%%%%%%%%%%%%%%%%%%%%%%%%%%%%%%%%%%%%%%%%%%%%%
\section{Results}\label{dec:results}
%%%%%%%%%%%%%%%%%%%%%%%%%%%%%%%%%%%%%%%%%%%%%%%%%%%%%%%%%%
\begin{figure}[t]
    \centering
    \includegraphics[width=0.45\linewidth]{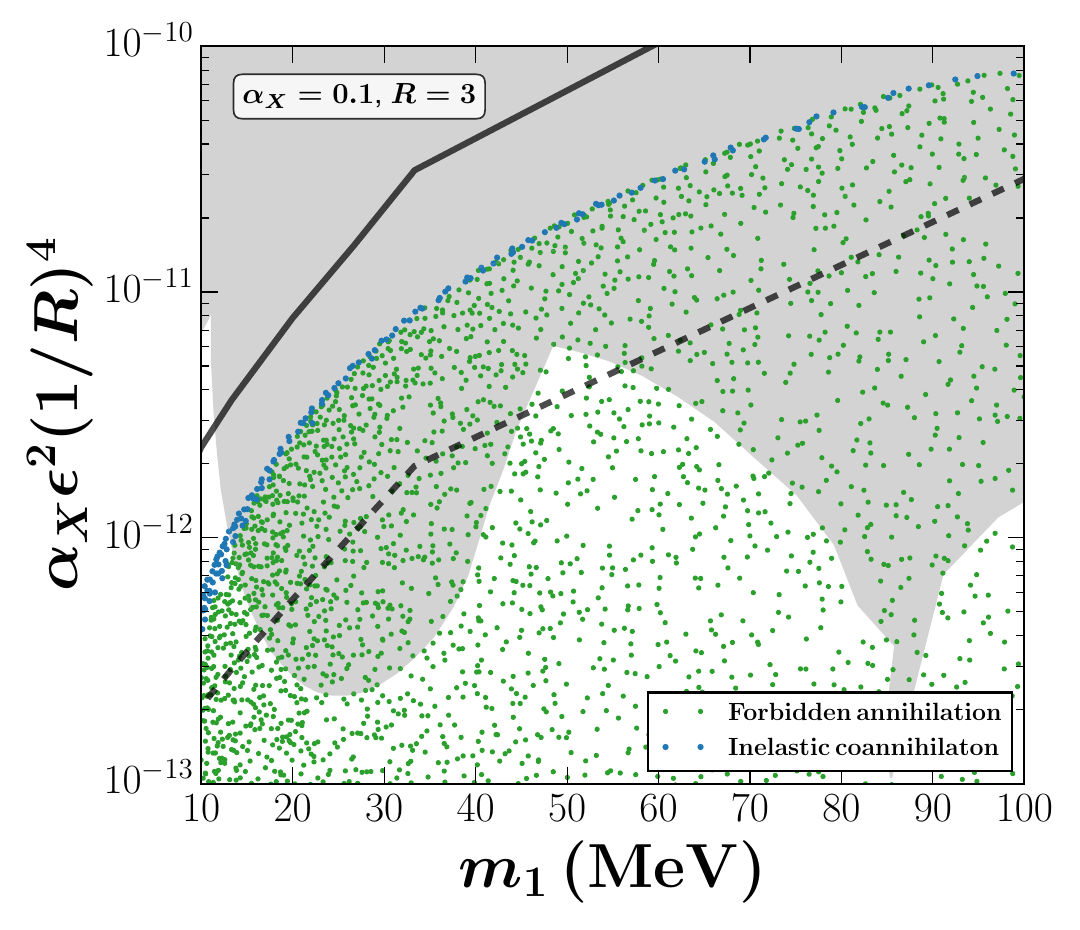}
    \includegraphics[width=0.45\linewidth]{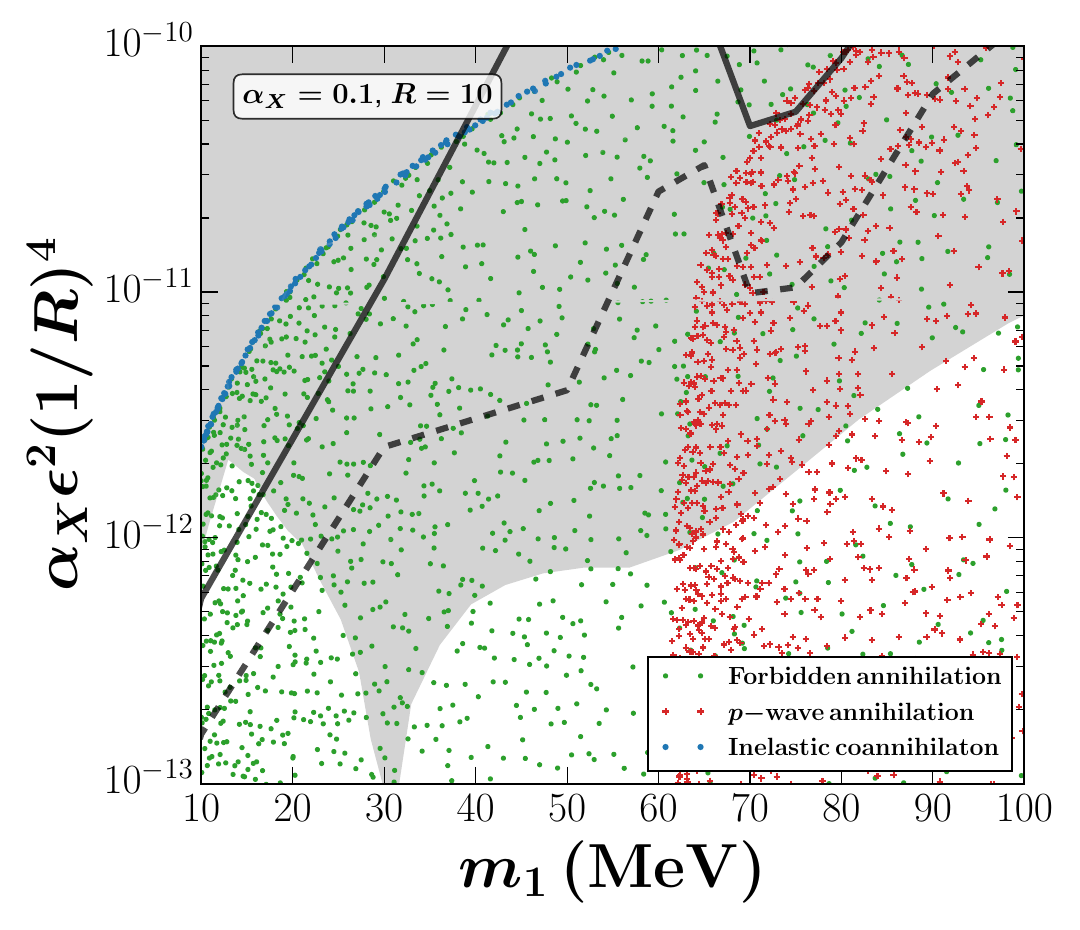}\\
    \includegraphics[width=0.45\linewidth]{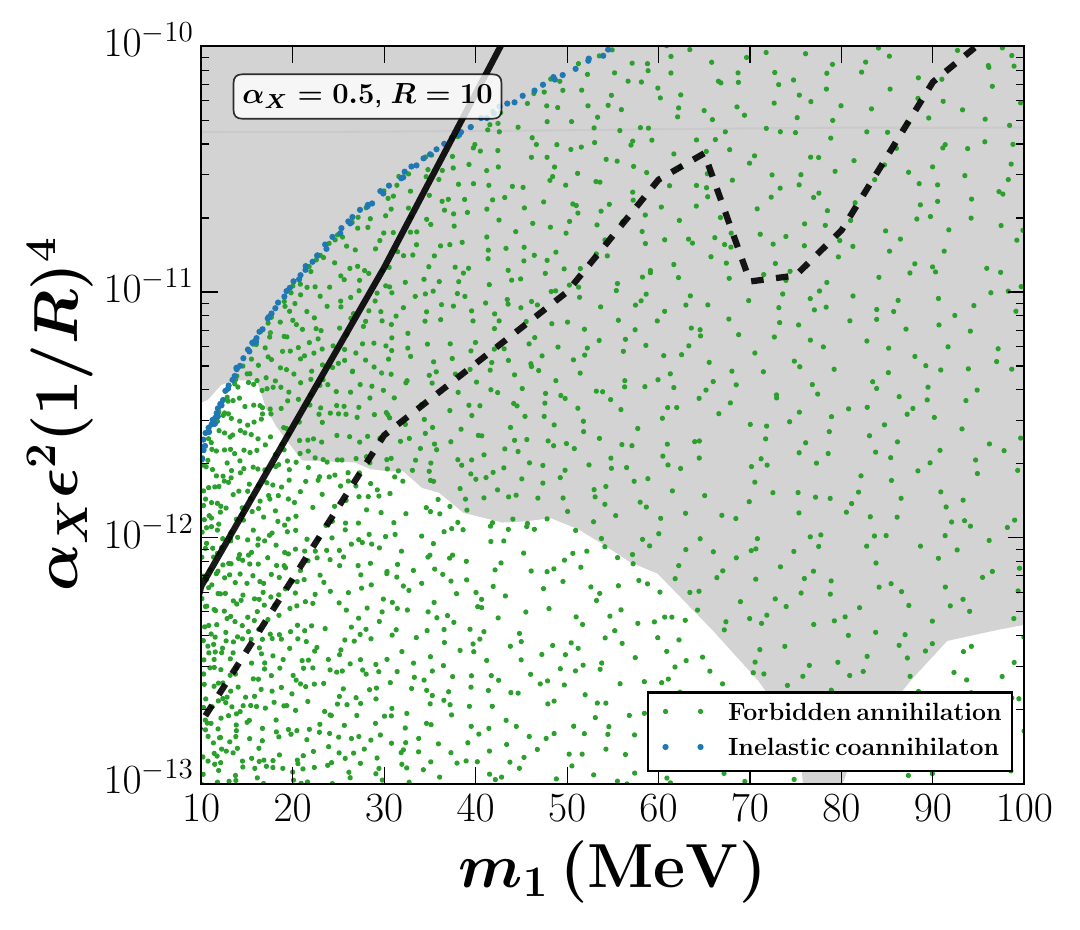}
    \caption{Parameter space in the plane $(m_1,y)$ reproducing the DM abundance for different choices of $\alpha_X$ and $\Delta$.  The black lines denote the DUNE ND-LAr sensitivity operating on-axis (solid) and in a beam-dump mode (dashed). 
    Green, blue, and red points correspond to regions where the relic abundance is determined predominantly by forbidden annihilation, $p$-wave annihilation, and inelastic coannihilation, respectively.
    The gray shaded regions are excluded by a combination of several experiments~\cite{NA64:2021acr,Mongillo:2023hbs}.  In all the panel $\sin\theta=5\times10^{-5}$, and $\Delta=0.1m_1$ are fixed. }
    \label{fig:DEscan2}
\end{figure}

Our main results are shown in the panels of Fig.~\ref{fig:DEscan2}. The dimensionless $y$ parameter defined as $\alpha_{X} \epsilon^2\left(\frac{m_{\chi_1}}{m_{A'}}\right)^4$ characterizes the cross-section of DM interaction with fermions mediated by a dark photon, which affects both relic abundance and fixed-target experiment sensitivity. For this reason, we present our results in the $y-m_1$ plane. We focus on 3 benchmark scenarios: $\alpha_{X}=0.1, R=3$, $\alpha_{X}=0.1, R=10$, $\alpha_{X}=0.5, R=10$, while keeping $\sin \theta=5\times10^{-5}$, and $\Delta=0.1 m_1$. We also show the current restrictions following Ref.~\cite{NA64:2021acr,Mongillo:2023hbs} from experiments such  BABAR and E137~\cite{Izaguirre:2017bqb, Mohlabeng:2019vrz}, NuCaI and CHARM~\cite{Tsai:2019buq}, and NA64 and BABAR~\cite{Mongillo:2023hbs} as shaded gray regions. We also note that the MiniBooNE beam-dump search provides an important existing constraint on light dark matter produced in proton beam dumps, especially with the inclusion of timing information ~\cite{MiniBooNE:2017nqe,MiniBooNEDM:2018cxm}. As such, it could, in principle, provide additional restrictions. Nevertheless, the published MiniBooNE results were obtained for complex scalar elastic DM, and therefore they cannot be directly placed on our parameter space without a dedicated recast.

All data points shown in Fig.~\ref{fig:DEscan2}. reproduce the observed relic abundance~\cite{Planck:2018vyg}.  To facilitate the discussion, we classify the viable points according to the dominant process responsible for setting the relic abundance. Green points correspond to the forbidden-annihilation regime, where the process $\chi_1\chi_1\rightarrow h'h'$ proceeds through the thermal tail because the final state is slightly heavier than the initial one ($m_{h'}\gtrsim m_1$). Blue points correspond to the kinematically open but $p$-wave--suppressed regime, in which the same annihilation channel remains dominant but is suppressed by the DM velocity at freeze-out. Finally, red points correspond to the standard inelastic coannihilation regime, where the relic abundance is determined primarily by the process $\chi_1\chi_2\rightarrow\bar{f}f$ mediated by the dark photon. The $p$-wave–suppressed region appears only in the top-right panel, since in the other cases it lies beyond the upper limit of the mass range shown.

For the case of $\alpha_{X}=0.1$ and $R=3$ (top left panel), the most widely studied benchmark, we find that DUNE's ND-LAr operating on-axis (solid black curve) has no sensitivity to regions of parameter space consistent with the observed DM relic abundance. However, if a dedicated beam-dump run (dashed black curve) is realized, DUNE could begin to probe points that achieve the observed relic abundance for $m_{1}\sim10$ MeV and $m_{1}\sim50$ MeV, although only at large enough values of $\epsilon$. 

When a larger mass ratio, $R=10$, is considered (top right panel), the correct relic abundance is achieved for higher values of the kinetic mixing parameter $\epsilon$. In this case, seven years of DUNE running on-axis demonstrate sensitivity to the region
$10~\mathrm{MeV} \lesssim m_1 \lesssim 15~\mathrm{MeV}$ with $y \sim 10^{-12}$. If the possibility of a beam-dump run is included, DUNE's reach is further extended, allowing sensitivity up to $m_{1} \sim 20~\mathrm{MeV}$ while probing values of $y$ lower than $10^{-12}$. The sensitivity prospects are slightly better if a large gauge coupling is considered (bottom panel). 

We want to point out two important points of our results: the inclusion of the forbidden channel significantly enlarges the region in the $y-m_1$ plane compatible with the observed relic abundance, while the pure inelastic regime sets an approximate upper envelope in $y$. The regions where $R$ is large are particularly challenging to explore with the usual search of $\chi_2$ decays; this is because, for this set of parameters, the heavier partner's stability is enhanced. In contrast, the ability of DUNE ND-LAr to probe DM-$e$ scattering enables sensitivity to this particular unexplored region. This highlights the relevance that DUNE can have in physics well beyond the reach of the neutrino sector.

\begin{figure}[t]
    \centering
    \includegraphics[width=0.45\linewidth]{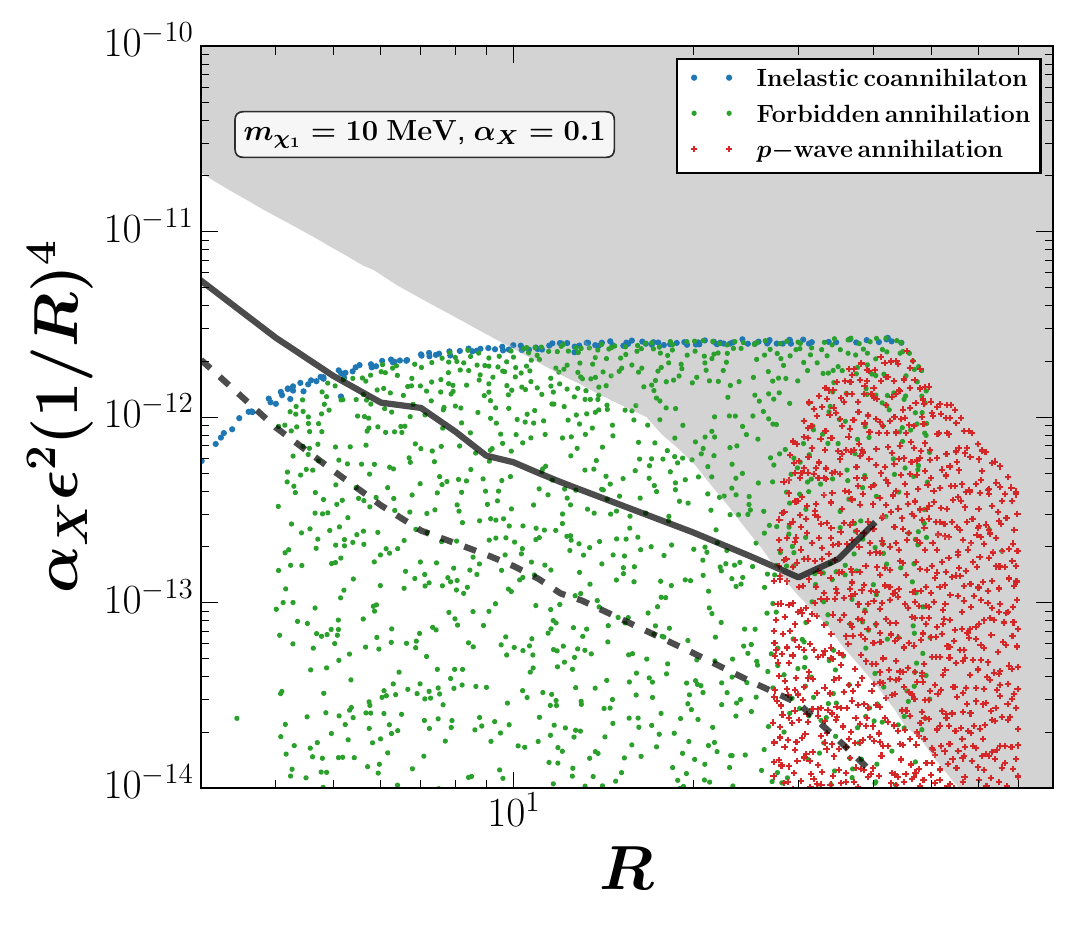}
    \caption{Parameter space in the plane $(R,y)$ reproducing the DM abundance for the benchmark scenario $\alpha_X=0.1$ and $m_1=10$ MeV.  The black lines denote the DUNE ND-LAr sensitivity operating on-axis (solid) and in a beam-dump mode (dashed). 
    Green, blue, and red points correspond to regions where the relic abundance is determined predominantly by forbidden annihilation, $p$-wave annihilation, and inelastic coannihilation, respectively.
    The gray shaded regions are excluded by NA64 and BABAR~\cite{Mongillo:2023hbs}.  $\sin\theta=5\times10^{-5}$ and $\Delta=0.1m_1$ have been fixed. }
    \label{fig:DEscan3}
\end{figure}

In Fig.~\ref{fig:DEscan3} we display the region of the parameter space that reproduces the observed relic abundance while fixing the DM mass and varying the mass ratio $R=m_{A'}/m_1$. The gray shaded areas correspond to existing constraints from NA64 and BABAR, as reported in Ref.~\cite{Mongillo:2023hbs}. This projection highlights how the viable thermal targets extend over a broad range of mediator-to-DM mass hierarchies once the dark Higgs--induced annihilation channels are taken into account. 

We find that both the standard on-axis DUNE configuration and the proposed beam-dump mode exhibit sensitivity to sizable regions of this parameter space, including values of $R$ significantly larger than the commonly studied benchmark $R=3$. This is particularly noteworthy, since large-$R$ scenarios are typically challenging for decay-based searches due to the enhanced stability of the heavier state and the reduced visible decay signatures. In contrast, DUNE's sensitivity to DM--electron scattering allows it to probe these otherwise difficult regions, thereby providing complementary coverage of thermal iDM scenarios with extended dark sectors.

%%%%%%%%%%%%%%%%%%%%%%%%%%%%%%%%%%%%%%%%%%%%%%%%%%%%%%%%%%
\section{Conclusions}\label{sec:conclusions}
%%%%%%%%%%%%%%%%%%%%%%%%%%%%%%%%%%%%%%%%%%%%%%%%%%%%%%%%%%
We have investigated an inelastic dark matter (DM) scenario featuring a Dirac fermion and a dark Higgs field as responsible for the spontaneous breaking of the $U(1)_D$ gauge symmetry. This symmetry breaking not only generates the dark photon mass but also naturally induces a mass splitting between two Majorana DM states, $\chi_1$ and $\chi_2$. As a result, new annihilation channels mediated by the dark Higgs arise, including forbidden annihilations and $p$-wave–suppressed processes, which play a central role in determining the DM relic abundance.

Focusing on the sub-GeV mass range, we have assessed the sensitivity of the DUNE ND-LAr to this scenario. We performed a detailed simulation of DM production and detection at DUNE, including realistic detector effects, backgrounds, and the finite lifetime of the heavier dark sector state. Our analysis shows that DUNE can probe regions of parameter space that are difficult to access with typical decay-based searches, particularly for large ratios of the dark photon to DM mass. 

This work represents the views and results of the authors and not those of the DUNE Collaboration as a whole.

%%%%%%%%%%%%%%%%%%%%%%%%%%%%%%%%%%%%%%%%%%%%%%%%%%%%%%%%%%
\section*{Acknowledgments}
%%%%%%%%%%%%%%%%%%%%%%%%%%%%%%%%%%%%%%%%%%%%%%%%%%%%%%%%%%
We are grateful to Camilo García-Cely for insightful discussions, and to Richard Diurba and Stefan Soldner-Rembold for valuable comments and suggestions.  We also thank the MadDump development team for helpful discussions and for their assistance with the implementation and use of the {\tt MadDump} package. O.Z. has been partially supported by Sostenibilidad-UdeA and the UdeA/CODI Grants 2022-52380 and  2024-76476. 
%%%%%%%%%%%%%%%%%%%%%%%%%
\bibliographystyle{JHEP}
\bibliography{references}
%%%%%%%%%%%%%%%%%%%%%%%%%
\end{document}